\documentstyle[11pt,aaspp4]{article}

\def\gapprox{\;\rlap{\lower 3.0pt                       % approximately larger
             \hbox{$\sim$}}\raise 2.5pt\hbox{$>$}\;}
\def\lapprox{\;\rlap{\lower 3.0pt                       % approximately smaller
             \hbox{$\sim$}}\raise 2.5pt\hbox{$<$}\;}

\begin{document}
\title{On the Surface Heating of Synchronously-Spinning Short-Period 
Jovian Planets}
\author{Andreas Burkert}
\affil{Max-Planck-Institut f\"ur Astronomie, Heidelberg, Germany \\ and \\ 
University Observatory Munich, Scheinerstr. 1, D-81679 Munich, Germany}
\author{Douglas N. C. Lin}
\affil{UCO/Lick Observatory, University of California, Santa Cruz,
CA 95064, U.S.A.}
\author{Peter H. Bodenheimer}
\affil{UCO/Lick Observatory, University of California, Santa Cruz,
CA 95064, U.S.A.}
\author{Chris A. Jones}
\affil{Department of Mathematical Sciences, Exeter University,  
Exeter, UK}
\author{Harold W. Yorke}
\affil{Jet Propulsion Laboratory, California Institute of Technology, 
 Pasadena, CA 91109, U.S.A.}
\centerline{\today}

\begin{abstract}
We consider the atmospheric flow on short-period extra-solar planets
through two-dimensional numerical simulations of hydrodynamics with
radiation transfer.  The observed low eccentricity of these planets
indicates that tidal dissipation within them has been effective in
circularizing their orbits and synchronizing their spins.
Consequently, one side of these planets (the day side) is always
exposed to the irradiation from the host star, whereas the other (the
night side) is always in shadow.  The temperature of the day side is
determined by the equilibrium which the planetary atmosphere
establishes with the stellar radiation.  For planets around solar-type
stars with periods less than 7 days, the flux of stellar irradiation
exceeds that released from their Kelvin-Helmholtz contraction by
several orders of magnitude.  A fraction of the thermal energy
deposited on the day side is advected to the night side by a current.
We show that the radiation transfer and the night-side temperature
distribution in a planet's atmosphere are sensitive functions of its
opacity.  If the atmosphere contains grains with an abundance and size
distribution comparable to that of the interstellar medium, only
shallow heating occurs on the day side, whereas the heat flux carried
by the circulation does not effectively heat the night side, which
cools well below the day side. The temperature difference affects the
spectroscopic signature of these planets.  However, the temperature
difference decreases as the abundance of grains in the atmosphere is
reduced.  This effect occurs because if the grains are depleted, the
stellar radiative flux penetrates more deeply into the atmosphere on
the day side, and the higher-density atmospheric circulation carries a
larger flux of heat over to the night side. A simple analytic  model of the
dissipation of the circulation flow and associated kinetic heating is also
considered. This heating effect occurs mostly near the photosphere,
not deep enough to significantly affect the size of planets.  The
depth of the energy deposition increases as the abundance of grains is
reduced.  Finally, we show that the surface irradiation suppresses
convection near the photospheric region on the day side.  But, in some
cases, depending on the opacity, convection zones are present near the
surface on the night side.  This structural modification may influence
the response and dissipation of tidal disturbances and alter the
circularization and synchronization time scales.

\end{abstract}

\section{Introduction}

One of the surprising findings in the search for extra-solar planets
was the discovery of short-period planets (Mayor \& Queloz 1995).
Tidal dissipation within the envelopes of planets in close proximity
to their host stars would induce them to spin synchronously with their
orbital frequency if their tidal dissipation function is comparable to
that of Jupiter (Yoder \& Peale 1981).  Consequently, one side (the
day side) of these planets would be continuously heated by the intense
stellar radiation while the other side (the night side) would be
permanently in shadow.

Besides its importance for the cooling of the planet, the temperature
distribution at the photosphere as a function of location on the
planet determines its luminosity as a function of orbital phase. This
property is especially interesting with regard to the direct
detectability of planets close to their stars.  It may also
significantly modify the spectroscopic signatures of these planets
(Sudarsky, Burrows, \& Pinto 2000; Seager \& Sasselov 2000; Brown
2001) from those computed under the assumption of uniform surface
temperature. The key issue to be addressed in this paper is the
photospheric temperature distribution around the surface of a close-in
giant planet and its dependence on the opacity in the planet's
atmosphere.

The response of the planetary atmosphere to the source of intense
stellar irradiation is important for determining other observable
properties of extrasolar planets.  For example, the planet around HD
209458 has been detected through transit observations (Charbonneau et
al. 2000; Henry et al. 2000).  The observationally inferred size of
this planet appears to be 1.35 to 1.42 times Jupiter's radius R$_J$
(Brown et al. 2001; Cody \& Sasselov 2002).  This relatively large
size of the planet is not consistent with that expected from the
evolutionary track of a gaseous giant planet with the appropriate
mass.  Bodenheimer, Laughlin, \& Lin (2003) and Baraffe et al. (2003)
find that the model radius is only about 1.1 R$_J$ at the estimated
age of HD 209458.  It has been suggested that heating of the outer
layers by stellar irradiation, which reduces the temperature gradient
and the radiative flux in those layers, could significantly slow down
the Kelvin-Helmholtz contraction of the planet and explain the large
size (Burrows et al.  2000). However, even though the stellar flux
onto the planet's surface is 5 orders of magnitude larger than that
released by the gravitational contraction and cooling of its envelope,
this heating effect alone increases the radius of the planet by about
10\%, not by 40\% as observed (Guillot \& Showman 2002).  An
additional energy source, beyond simple stellar irradiation,
contraction, and internal cooling, is needed.

A possible energy source is tidal dissipation within the planetary
envelope (Bodenheimer, Lin, \& Mardling 2001), caused by the
circularization of the orbit.  The tidal heating scenario can be
tested observationally because it requires a small but non-zero
eccentricity for the planet ($e \approx .03$) which needs to be
induced by another hypothetical planet (Bodenheimer et al.  2001).
While current observations suggest that both of these requirements
could be satisfied (Bodenheimer et al.  2003), the data are not
sufficiently accurate to reach any firm conclusions.

An alternative source is the kinetic heating induced by the
dissipation of the gas flow in the atmosphere which occurs because of
the pressure gradient between the day and night sides (Guillot \&
Showman 2002).  In order to account for the observed size of the
planet, conversion of only 1\% of the incident radiative flux may be
needed, provided that the dissipation of induced kinetic energy into
heat occurs at sufficiently deep layers (tens to 100 bars).  Showman
\& Guillot (2002) suggest that the Coriolis force associated with a
synchronously spinning planet may induce the circulation to penetrate
that far into the planet's interior and that dissipation could occur
through, for example, Kelvin-Helmholtz instability.  A follow-up
analysis suggests that this effect may be limited (Jones \& Lin 2003).
The kinetic heating scenario requires a demonstration, which has not
been carried out yet, that there exists a sufficient dissipation
mechanism. It would depend on the details of the circulation flow in
the atmosphere.  In this paper we analyze that flow and suggest a
simple alternative model for kinetic heating: viscous heating arising
as a consequence of shear flows that develop in an atmosphere.  We
consider the question of whether that heating is sufficient to result
in expansion of the planet.

The three-dimensional numerical simulation presented by Showman \&
Guillot (2002) was designed to strengthen the case for a dissipation
mechanism in the deeper layers of the atmosphere.  Using approximate
local heating and cooling functions, they show that the kinetic energy
flux transported downward over the 100 bar surface, induced by the
differential heating from the star, amounts to about 1\% of the
absorbed stellar flux. It is plausible that this energy is dissipated
at the deeper layers by, for example, the Kelvin-Helmholtz
instability, but it has not been shown how efficient this mechanism
is.  In the same paper the authors estimate the day-night temperature
difference (their equation 11):
\begin{equation}
{{\Delta T_{dn}} \over {\Delta T_{rad}} } \sim 1 - \exp (-\tau_{\rm zonal} /
\tau_{\rm rad}),
\label{eq:dn}
\end{equation} 
where $\Delta T_{rad}$ is the difference in radiative equilibrium
temperature between the day side and the night side, $\tau_{\rm
zonal}$ is a characteristic time for winds to advect energy from the
day side to the night side, and $\tau_{\rm rad}$ is the radiation
diffusion time.  This formula, with reasonable estimates for the time
scales, indicates that $\Delta T_{dn}$ could be $\sim 500$ K near the
photosphere (at $\sim 1$ bar) for a close-in giant planet. The value
of $\Delta T_{dn}$ is thus strongly dependent on what is assumed for
$\Delta T_{rad}$.  The numerical simulations do not show such a large
temperature variation simply because $\Delta T_{rad}$ was chosen to be
only 100 K, leading to $\Delta T_{dn} \approx 50$ K, in reasonable
agreement with equation (\ref{eq:dn}). However, a tidally locked planet
would be expected to have a night-side temperature of $\sim 100$~K
from its internal cooling, and a day-side temperature of at least 1000
K.  The simulations do not include radiation transport, and especially
if the effect of stellar irradiation leads to flows at high optical
depths, the effects of radiation diffusion could be quite important in
the analysis of the kinetic energy flow in the deep envelope.  In
contrast, a two-zone analytic approximation by Jones \& Lin (2003)
suggests, for a standard grain opacity, a more substantial ($\sim 500$~K)
mean temperature difference between the day and night sides of the
planet.
 
Cho et al. (2003) perform a two-dimensional hydrodynamic calculation
of the flows across the surface of a close-in Jovian planet, using the
shallow layer approximation to integrate over the vertical direction.
The heating rate is parameterized and radiation transfer is not
included.  In the case that is presented in the paper, the mean
day-to-night temperature variation is assumed to be 180 K, but the
calculations show that, superimposed on this background variation,
hot-spot/cold-spot pairs appear  close to each pole, which
circulate about the poles. The maximum temperature variation between
 the hot spot and the  cold spot is about 300 K, but it could be larger
depending on parameters.  In both sets of simulations the local
cooling approximation is a gross simplification which greatly
affects  the outcome of the flow. Since the flow is driven by the
thermal properties of the atmosphere, it is clear that more detailed
numerical models with improved physics are required to understand the
complex, nonlinear radiation hydrodynamics.  The validity of these
simulations may be potentially tested by direct observations of the
temperature difference between the day and night sides of short-period
planets.

In the absence of any significant surface irradiation, the envelope of
a mature gas giant planet is nearly fully convective.  Surface
irradiation modifies the radial temperature gradient and suppresses
convection (Bodenheimer et al. 2001) at least in the upper regions on
the day side.  In the radiative regions of the planet's envelope, g
modes (Ioannou \& Lindzen 1993) and Hough modes (Ogilvie \& Lin 2003)
are dynamically excited by the host star's tidal disturbance.  The
amplitude and dissipation rate of the gravity and inertial waves
determine the planet's Q values and circularization and
synchronization time scales (Goldreich \& Nicholson 1989; Lubow et
al. 1997).  Thus, it is also useful to evaluate the extent of the
radiative zone near the planet's surface.

In this paper we carry out two-dimensional hydrodynamic simulations,
incorporating  a flux-limited diffusion treatment of radiation transport,
of the flow pattern in the envelope of a synchronously spinning
extra-solar Jupiter-like planet which is heated by a close-by star.
The two dimensions are the depth and the azimuthal position on the
surface.  The quantity $\Delta T_{\rm rad}$ is assumed to have the
realistic value of 1100 K, but there is no built-in assumption
regarding $\Delta T_{\rm dn}$.  For computational simplicity we first
consider the case in which the effect of rotation is limited.  
The spin period of a synchronously rotating close-in planet is an
order of magnitude longer than that of Jupiter or Saturn.  The Rossby
radius $v/\Omega$ where $v$ is a typical flow velocity (a few km
s$^{-1}$) and $\Omega$ is the planetary angular frequency, is
therefore a few times larger than the radius of the planet so the
Coriolis effect will not be dominant, but it may be of some
importance.  We therefore plan to consider the effect of rotation in a
future investigation, but focus in this paper only on the influence of
radiation transport on the dynamical structure of the planet's
atmosphere.

In \S2, we briefly recapitulate the basic equations for mass,
momentum, and heat transfer and describe a 2-D numerical radiative
hydrodynamic scheme with which we compute the flow and structure in
the planetary envelope on both the day and night side.  We also
explicitly state our basic assumptions and parameters and specify our
initial and boundary conditions.  In \S3 we describe the results of
our standard model, which is based on the opacity of the interstellar
medium, and additional models in which we vary the opacity by 5 orders
of magnitude.  We show that the distribution of effective temperature
over the surface is determined by the advective heat flux close to the
photosphere, and that the night-side temperature depends mainly on the
opacity of the photosphere on the day side. In \S4 we estimate the
extent of kinetic heating induced by the calculated velocity shear and
an assumed viscosity.  Finally, in \S5 we summarize our results and
discuss some of their implications.

\section{The Computational Method}

\subsection{The basic equations}

In order to investigate the gas flow in the planetary atmosphere, we
have performed 2-dimensional hydrodynamical calculations adopting an
Eulerian, uniform Cartesian grid code with radiation transport to
treat properly stellar heating and radiative cooling. In this
approximation we neglect the effect of latitudinal motion, as the
rotational effect is assumed to be small.  Since the vertical extent
of the interesting surface region is small compared with the radius of
the planet, geometrical curvature is negligibly small (see \S 2.2).
The hydrodynamical equations are:

\begin{eqnarray}
\frac{\partial \rho}{\partial t} + \vec{\nabla} \cdot (\rho \vec{v}) =0 \\ 
\frac{\partial(\rho \vec{v})}{\partial t} + \vec{\nabla} (\rho
\vec{v} \vec{v}) = - \vec{\nabla} P - \rho \vec{g} \\ \frac{\partial
(\rho e)}{\partial t} + \vec{\nabla}\cdot (\rho e \vec{v}) + P
 \vec{\nabla}\cdot \vec{v}=-\vec{\nabla} \cdot \vec{F},
\label{eq:energy} 
\end{eqnarray}
\noindent where $\rho$, $e$, $P$ and $\vec{v}$ are the gas density,
internal energy per unit mass, pressure and velocity, respectively.
In the atmosphere the acceleration of gravity $\vec{g}$ is assumed to
be constant.  The radiative flux $\vec{F}$ is calculated according to
flux-limited radiative diffusion
\begin{equation}
{\vec{F} }= -{ {c \lambda } \over { \kappa_R \rho}}{\vec{\nabla}}{u}, 
\label{eq:limit}
\end{equation}
where $c$ is the speed of light, $\lambda$ is the flux limiter, $u$
is the black-body radiation energy density $u = aT^4$, where $a$ is
the radiation density constant, and $ \kappa_R$ is the Rosseland mean
opacity.  The flux limiter is approximated according to the procedure of
Levermore \&  Pomraning (1981):
\begin{equation}
\lambda = { {2 + R} \over {6 + 3R + R^2}}, 
\label{eq:2R}
\end{equation}
where $R$ is the ratio of the mean free path of a photon to the scale
height of the radiation  energy density:
\begin{equation}
R = { {\vert{\bf \nabla} u \vert} \over {\kappa_R  \rho  u}} . 
\end{equation}
Note that in the optically thick limit $R \rightarrow 0$ and equation
(\ref{eq:limit}) reduces to the normal diffusion equation for stellar
interiors, and in the optically thin limit it reduces to the 
free-streaming solution  $\vert F \vert \rightarrow
cu$.

The main sources of opacity in the atmosphere of the planet, which
ranges typically from 100 K to 1500 K in temperature, are ice-coated
silicate grains at low T, and grains composed of silicates, amorphous
carbon, and iron at temperatures above the ice evaporation temperature
which is of order 170 K. The Rosseland mean opacities are taken from
Pollack, McKay, \& Christofferson (1985) and Alexander \& Ferguson
(1994), who assumed an interstellar size distribution and solar
metallicity for the grains. Because 1) the size distribution in the
planet's atmosphere can be different from that in the interstellar
medium, 2) the grains in the upper atmosphere can be depleted due to
their sedimentation to layers where they can evaporate, and 3) the
planetary envelope may be enriched in heavy elements by an order of
magnitude relative to their host stars, we also considered models in
which the grain opacity is arbitrarily reduced or increased from the
standard values by a factor of $f_{\kappa}$ in the range of $10^{-3} -
10^2$.  The high-$f_\kappa$ values represent grain-enriched
atmospheres whereas the low-$f_\kappa$ values represent grain
depletion.

\subsection{The numerical method}

The full computational region consists of a rectangular area with size
$x_{max} \times (y_{max} - y_{min}) $ and $NX \times NY$ grid cells,
equally spaced in both directions. The $x$ and $y$ coordinates
represent the azimuthal and radial directions, respectively.  The
point $ y_{min}$ corresponds to the base of the atmosphere, and $
y_{max}$ represents the outer edge of the grid, which is well above
the level of optical depth $\tau = 1$.  The $x$ direction corresponds
to a strip around the circumference of the planet, with $x_{min} = 0$
at the sub-stellar point and $x_{max} = \pi \cdot y_{min},~180^\circ$
away on the cold side of the planet. Since the depth of the
atmosphere, $(y_{max} - y_{min})$, is very small compared with
$x_{max}$, the effects of curvature are negligible.  In a few cases we
tested the assumption of semi-cylindrical symmetry by using an x-grid
extending the full 360$^\circ$ around the planet. For the non-rotating
cases presented here, we found no significant signs of symmetry
breaking.  The set of differential equations is integrated by means of
an explicit finite difference scheme with operator splitting as
described by Burkert \& Bodenheimer (1993, 1996). The advection terms
are treated using the second-order monotonic transport scheme proposed
by van Leer (1977).  An artificial viscosity of the type as described
by Colella \& Woodward (1984) is added in order to suppress numerical
instabilities.  The time rate of change of internal energy caused by
radiation transport is treated in a separate radiation substep, which
is carried out implicitly. The two-dimensional diffusion equation is
solved iteratively using the alternating-direction implicit technique
(Press et al. 1992). The radiation transport module is very similar to
that used in protostellar collapse simulations by Yorke \& Bodenheimer
(1999).

The boundary conditions for the hydrodynamics are as follows: at each
of the four boundaries the perpendicular component of the velocity is
set to zero, as are the gradients, perpendicular to the surface, of
the density, internal energy, and the velocity component parallel to
the surface.  The boundary condition for the radiation transfer at the
inner boundary of the atmosphere ($y = y_{min}$) is a constant
temperature (in time and as a function of $x$).  The outer boundary
condition is based on the assumption that a mature planet sufficiently
close to the star rotates synchronously with its orbital motion
(assumed circular). The side facing the star is heated to a
temperature of $T_{equ} \cdot (\cos \theta)^{1/4}$, where $\theta$
is the angle between the directions of the host star and the normal to
the surface and $T_{equ}$ is the equilibrium temperature of the
planetary surface in the radiation field of the star. At angles where
this temperature falls below 100 K, which includes the entire night
side of the planet, it is reset to 100 K, roughly the value expected
from the internal cooling of an isolated planet.  This nominal
temperature on the night side corresponds to a very low-density upper
atmosphere, where the optical depth falls well below unity, and its
precise value does not affect the solution as long as it is low
compared to the temperature on the day side.

\subsection{Initial conditions}

The standard model is motivated by the observations of the transiting
planet HD 209458b. The model planet has solar composition, no rock/ice
core, a mass of 0.63 Jupiter masses (M$_J$), and an outer radius
($R_p$) of 1.2 Jupiter radii (R$_J$).  It is the end result, at a time
of 4.1 Gyr, of an evolutionary calculation starting at a radius of 2
R$_J$ for a spherically symmetric planet assumed to be in orbit at
0.046 AU from a solar-type star (model B1 of Bodenheimer et al. 2001).
The outer region, down to a density of 0.1 g cm$^{-3}$ or 88.5\% of
the outer radius, is radiative in the spherical model.  The grain
opacities are the same as those used for the two-dimensional
calculations for Case 1 (see below) and are based on the interstellar
particle size distribution with solar composition.  The stellar
heating gives $T_{equ} = 1200 $ K.

The 2-D grid is chosen to have its lower boundary at $y_{ min} = 7.95
\times 10^9$ cm and its upper boundary at $y_{max} = 8.35 \times
10^9$ cm.  All calculations were performed on a grid with $NX=150$ and
$NY=50$ grid cells in the azimuthal and radial directions,
respectively.  The spatial resolution is therefore $ \Delta y = 8 \times 10^6$ cm
and $\Delta x = 1.6 \times 10^8$ cm. 
In the spherical model $y_{min}$ corresponds to a temperature of 1430
K and a density of $10^{-4}$ g cm$^{-3}$.  The intrinsic luminosity of
the planet, without the effects of stellar heating, is $L_p = 2.6
\times 10^{24}$ erg s$^{-1}$. Initially, the density and temperature in
the 2-D grid, $\rho(y)$ and $T(y)$, are set to values similar to those
in the one-dimensional model for each value of $x$.  The temperature
of 1430 K, as found in the spherically symmetric calculation at
$y_{min}$, is adopted as the inner boundary condition for the
radiation transfer in all cases.

With the $\theta$-dependent new temperature boundary condition turned
on, the spherically symmetric structure is lost owing to cooling on
the night side, and a gas flow between day and night side is
generated. The two-dimensional calculation is evolved with an initial
damping of the gas flow velocities of 1\% per time step, until the
configuration has adjusted to approximate hydrostatic equilibrium. 
Then the damping is turned off, and the model is evolved for a few
$10^7$~s.  This computational time corresponds to several hundred
times the characteristic flow time $\tau_{\rm zonal}\approx
R_p/v\approx 10^5$ s.  To compare with the local cooling time scale
\begin{equation} 
 \tau_c \sim \Sigma R_g T /F, 
\label{eq:tc}
\end{equation}
where $R_g$ is the gas constant, 
we note that the surface density above the planet's photosphere is
$\Sigma_p \sim \kappa^{-1}$ (see equation \ref{eq:pd} below).  In the
normal-$\kappa$ ({\it i.e.}  $f_\kappa=1$) model the cooling time at
a mature planet's photosphere is $\approx 200$ s, as the relevant flux
to be used in equation (\ref{eq:tc}) is the radiative flux arriving
from the star.  In the low-$\kappa$ ({\it i.e.}  $f_\kappa=10^{-3}$)
model the cooling times are a factor of 1000 longer but still short
compared to the computation time.  Also, if we use the expression
given for $\tau_{\rm rad}$ (their equation 10) by Showman \& Guillot
(2002), the cooling time is on the order of 10$^6$ s or less in the
physical regime we are simulating.  In this case the computational
time is still many times greater than the radiation cooling time.

However, if one applies equation (\ref{eq:tc}) to the deeper layers of
the atmosphere, where $\rho \approx 10^{-5} - 10^{-4}$ g cm$^{-3}$,
and uses the intrinsic flux, one finds that $\tau_c \sim 10^{10}
f_\kappa^{-1}$ s, which is longer than the physical time scales we
have computed in our simulations.  The deeper layers of the models in
principle have not had time to reach a fully relaxed thermal
equilibrium.  But in fact the standard case $f_\kappa=1$ is started
from a thermally relaxed initial condition which was generated by a
spherically symmetric heated 1-D model.  Thus, the base of the
computational domain is close to a steady state.  In the cases where
$f_\kappa \ne 1$ radiative equilibrium is actually not achieved in the
very deep layers, but this does not affect our results for the main
part of the flow, which is dominated by the upper layers which are
influenced entirely by the stellar flux, 5 orders of magnitude larger
than the intrinsic flux.

\subsection{Computational model parameters}

In the standard model (Case 1) we consider a planetary atmosphere (1)
with a grain opacity which is identical to that of the interstellar
medium with solar composition and (2) with an equilibrium temperature
at the planet's surface of 1200 K (see Table 1). Regarding the
opacity, short-period planets have masses comparable to that of
Jupiter.  If their composition is similar to that of Jupiter's
interior (Wuchterl, Guillot, \& Lissauer 2000), which is
metal-enhanced compared with solar abundances, the atmospheric opacity
may be larger than that adopted in the standard model. To test the
limits we consider an extreme case in which the opacity is enhanced by
a factor of 100 with respect to the standard model (Case 2).
Alternatively, the grains are subject to settling, coagulation, and
evaporation in deeper layers, so their opacity could well be less than
that used in the standard model. We consider reductions in the
standard opacity by factors of 100 (Case 3) and 1000 (Case 4).

\section{The effect of opacity on the atmospheric flow 
and the day-night temperature difference}

\subsection{The standard case: interstellar opacity}

With an unmodified opacity the photosphere of the planet on both the
day and night sides extends to $\sim 3 \times 10^8$ cm above the base
of our computational domain.  The first frame of Figure 1 shows the
temperature distribution and the velocity field after steady state has
been reached. The inner regions at the base of the computational
domain are not affected by the heating and cooling of the outer
layers. Their temperature distribution is independent of $x$.  On the
other hand, near the terminator ($\theta = 90^\circ$) a strong
horizontal temperature gradient is established at the layers near the
photosphere where the optical depth is $\tau = 2/3$ (thick solid
line), owing to stellar heating on the day side and radiative cooling
on the night side. This temperature gradient, in combination with a
density gradient in the $x$-direction for a given $y$, induces a
strong horizontal pressure gradient.  In Figure 2 we plot $f_p = P
/P_{\rm av}$, where $P_{\rm av}$ is the pressure averaged over all
values of $x$ for a given value of $y$.

The pressure gradient generates a gas flow toward the night side both
above and somewhat below the photosphere, transporting heat and
resulting in a temperature at the photosphere on the night side of
about 500 K.  At the anti-stellar point ($\theta = 180^\circ$),  where
the circulation currents from the two hemispheres converge, a pressure
ridge is established.  The location of this ridge results from the
hemispherical symmetry of the system, which might be broken if
Coriolis forces are taken into account.  This pressure ridge drives a
return flow of cool gas from the night side toward the day side below
the photosphere at optical depths $\tau \approx 50$. This flow
generates a cool layer with $T \approx 900$ K  on the day side
between the externally heated outer layers ($T \approx 1200$ K) and
the internally heated inner layers ($T \approx 1400$ K).

In the first panel of Figure 3 we plot the local rate of change of
internal energy per unit volume caused by radiation. In order to
identify the locations where heating and cooling are most efficient,
we only consider regions where $\rho > 10^{-9}$ g cm$^{-3}$ and $\tau
> 10^{-2}$, where radiative flux is transported mainly by diffusion.
In Figure 1 we showed that the temperature near the day-side
photosphere is heated to a local maximum.  Below the photosphere the
envelope's temperature decreases inward as a result of the advection
of cooler material from the night side in the return flow. The inward
temperature gradient leads to strong radiative heating, which
compensates for the advective cooling and in addition generates an
expansion near the axis $\theta=0$, which drives an upward flow.  This
effect completes the circulation loop in the $(x,y)$ plane.

 In the upper layers near the terminator ($\theta = 90^\circ$) warm
material moves horizontally into the shadowed region and cools
strongly.  On the night side the intensity of this radiative heat
loss is a decreasing function of the azimuthal distance from the
terminator, since the horizontally moving material at angles $\theta
>> 90^\circ$ has already cooled strongly.  In thermal equilibrium the
loss of internal energy by radiation in this region is replenished by
both the advection of excess thermal energy and the compressional PdV
work.  In the region of the pressure ridge near $\theta = 180^\circ$,
the kinetic energy that is converted to thermal energy through PdV
work is in part radiated away, heating the layers above and below.
But because the region is optically thick and the radiation loss is
inefficient, the temperature there remains slightly higher than that
in the surroundings.  The corresponding pressure gradient is able to
drive the return flow.  In much of the grid, however, the radiative
contribution is quite small, and we find that the advective energy
transport and PdV work compensate each other (eq. \ref{eq:energy}).

\subsection{An analytic model of the thermal current}
In Table 1 we summarize some important physical values at the
photosphere on the day and night side. With the standard opacity the
photosphere pressure on the day side is $P_d \approx 1 \times 10^{-3}$
bar, which is consistent with the condition for hydrostatic
equilibrium,
\begin{equation}
P_d = {2 g \over 3 \kappa_d},
\label{eq:pd}
\end{equation}
where $g = GM_p/R_p^2 \sim 10^3$cm s$^{-2}$ is the planet's gravity,
$M_p$ is its mass, $R_p$ is its radius at $\tau = 2/3 $, and $\kappa_d
\approx 1 $ cm$^2$ g$^{-1}$ is the opacity on the day side at the
photosphere.  The gas temperature in the outer layers on the day side
is $T_d \simeq 1200$K with a corresponding sound speed $c_d \approx
2.6 \times 10^5$ cm~s$^{-1}$. The photospheric density on the day side
is then $\rho_d = P_d \mu /(R_g T_d) \approx 3 \times 10^{-8} $g
cm$^{-3}$ (where $\mu$ is the mean molecular weight), 
and the density scale height is
\begin{equation}
h_d = c_d^2/g \approx  6 \times 10^7 {\rm cm}. 
\end{equation} 
Note that as long as the temperature is greater than 400 K, $h_d$ is
larger than the vertical grid spacing. However, for lower T the scale
height is not well resolved.  (The photospheric densities and
temperatures in Case 2 on the night side therefore may be
under-resolved.)  The total thermal energy per unit area retained in
this exposed layer is
\begin{equation} 
E_h \sim \rho_d h_d c_d^2 \sim P_d h_d \sim {2 c_d^2 / (3 \kappa_d}).
\end{equation}
When this heated layer is advected to the night side with a speed
$v_d$, it carries an excess flux
\begin{equation}
F_a \sim {\rho_d v_d R_g (T_d-T_n) \over \mu} .
\end{equation}
In Case 1 $F_a \sim 2 \times 10^8$ erg s$^{-1}$ cm$^{-2}$, which is
comparable to the incident flux due to the stellar irradiation and is
much larger than the radiative flux due to the planet's internal
cooling, $F_p = L_p / (4 \pi R_p^2) \approx 3 \times 10^3$ erg
s$^{-1}$ cm$^{-2}$.  Thus, the surface layer of the night side is
primarily heated by the advective transport.

As the circulation current reaches the night side, it cools on a
characteristic time scale
\begin{equation}
\tau_c \simeq {E_h \over F_n} = {E_h \over \sigma T_n^4},
\label{eq:tauc}
\end{equation}
where $F_n$ is the radiative flux on the night side and $T_n$ is the
effective temperature at the photosphere of the night side. The night
side temperature is determined by the condition that 
\begin{equation}
\tau_c \sim \tau_x = \pi R_p / (2 v_d),
\end{equation}
where $\tau_x$ is the flow crossing time scale.  As long as this
equality gives $T_d > T_n$, the equilibrium is expected to be
established and to be self regulated, because (i) if $\tau_x > \tau_c$,
the gas temperature on the night side would decline, leading to a
larger pressure gradient, shorter $\tau_x$ and a longer $\tau_c$; or
(ii) if $\tau_x < \tau_c$, the cool side would tend to heat up, thus
reducing the driving pressure differential and forcing $\tau_x$ up,
while $\tau_c$ decreases.

The upper left panel of Figure 4 shows the temperature at the
photosphere from day side to night side for the standard model, which
varies from a maximum of 1200 K to a minimum of 480~K. The upper right
panel shows the $x$ component of the flow velocity close to the
photosphere as a function of $x$: on the night side it maintains a
nearly constant value of about 3.5 $\times 10^5$ cm s$^{-1}$, which,
as expected, is close to $c_d$ at 1200 K.

The advection speed $v_d$ is induced by the pressure difference
between equipotential surfaces on the two sides of the planet (Showman
\& Guillot 2002; Jones \& Lin 2003).  The maximum magnitude for $v_d$
is of  the order of  the sound speed on the day side $c_d$.  However, in a
thermal equilibrium with a relatively small $F_p$ the magnitude of
$v_d$ is also limited by the condition that the {\it net} rate of
energy that is transported from the day to the night side by advection
\begin{equation}
\dot E_a \sim \int \frac{\rho v_d R_g 
(T_d-T_n)}{\mu} 2 \pi r dr
\label{eq:edota}
\end{equation}
must not exceed the stellar energy input rate        
\begin{equation}
{\dot E}_d \simeq \pi R_p^2 F_d
\label{eq:eddot}
\end{equation} 
on the day side. The integral in equation (\ref{eq:edota}) extends
over the vertical distance corresponding to the day-night flow. Here
$F_d = L_\ast / 4 \pi a^2$ is the stellar irradiation flux with
$L_\ast$ the luminosity of the host star and $a$ the semimajor axis
of the planet's orbit. The condition ${\dot E}_a \leq {\dot E}_d$
leads to
\begin{equation}
{v_d \over c_d} \le \left( {3 \kappa_d R_p \sigma T_d^4 \over 4 c_d^3} 
\right) \left( {T_d \over T_d - T_n } \right),
\end{equation} 
which for the standard case is satisfied even when $v_d =c_d$.  The
hot regions on the day side cool efficiently when crossing the
terminator due to their low heat capacity and collapse into a dense
cold layer on the night side, where they lead to a relatively small
photospheric temperature. Under certain circumstances, however, the
{\it gross} rate of thermal energy transported from the day to the
night side in the upper atmosphere
\begin{equation}
\dot E_g \sim \int \frac{\rho v_d R_g 
T_d}{\mu} 2 \pi r dr
\end{equation}
could be much larger than ${\dot E}_d$ -- the limits are the same as in
equation (\ref{eq:edota}).  In this limit $T_n \approx
T_d$ and the heat advected in the optically thin layer is closely
balanced by that carried in a return flow well below the photosphere,
resulting in little {\it net} heat transport per circulation.

On the night side the thermal equilibrium condition ($\tau_c \sim
\tau_x$) is satisfied when
\begin{equation}
T_n \approx \left({4 v_d c_d^2 \over 3 \pi \kappa_d \sigma R_p}
\right)^{1/4},
\label{eq:tn}
\end{equation}
which is valid if $T_n < T_d$. For the standard model $T_n \approx
490$~K, which is fairly consistent with the numerical result of 480 K
(Fig. 4).  Note that $T_n$ is not affected by the night side's opacity, 
because it radiates efficiently under the black-body law.  This
approximation is adequate provided the advective heat flux $F_g = \rho
v_d R_g T_d/\mu$ is larger than the radiative flux emerging from the
planet's interior. However, for low values of $\kappa_d$ the total rate
of thermal energy transport that is carried by the flow is large even
though $T_d - T_n$ decreases, because $\rho_d \sim 1/\kappa$. As soon
as $F_g$ becomes large compared to the radiated flux, $T_n \approx
T_d$. We will test this prediction in the next section.

We now consider qualitatively the effect of stellar irradiation on the
process of heat transport from the planet's interior.  Burrows  et
al. (2000) suggested, on the basis of a spherical model, that surface
heating reduces the magnitude of the temperature gradient near the
surface, deepening the surface radiative zone and decreasing the
energy loss rate and contraction rate, as compared with an isolated
planet.  We find that the temperature on the night side can be
significantly smaller than that on the day side, depending on the
opacity. Thus,  the temperature gradient is increased on the night side
relative to the day side, increasing the radiated flow from the
interior. Heated planets with high-opacity atmospheres can therefore
cool and contract faster than the spherical approximation would lead
one to believe, but still not as fast as unheated, isolated planets.
However, in the limit where $f_\kappa \rightarrow 10^{-3}$ the temperature
distribution is almost spherically symmetric (see below and
eq. \ref{eq:tn}), so that the approximation of Burrows et al. (2000, 2003)
would apply. Note that the enhancement of planetary size by this
effect is now thought to be smaller than the work of Burrows et
al. (2000) indicated (see Guillot \& Showman 2002).  It would be
interesting in further work to calculate this effect quantitatively by
combining a long-term spherically symmetric evolution with a
two-dimensional atmosphere as a boundary condition.  The atmosphere
would have to go deep enough so that the temperature and density are
practically constant with longitude and would have to be relaxed to
thermal equilibrium everywhere.

\subsection{Cases with modified opacities}

Table 1 compares significant properties of the various numerical
simulations with different opacities at the photosphere. The
quantities $T_n^n$, $\rho_n$, and $P_n$ refer to the temperature (in
K) , the density (in g cm$^{-3}$), and pressure (in bar) at the
photosphere on the night side.  (A superscript `$n$' is added to $T_n$
in column 2 to represent the value obtained directly from numerical
simulations.)  The corresponding quantities with subscript `d' refer
to the day side. The quantity $\kappa_d$ refers to the Rosseland mean
opacity on the day side at the photosphere, in cm$^2$ g$^{-1}$, while
$v_{max}$ is the maximum horizontal flow velocity from the day side to
the night side at the photosphere, which usually occurs over a broad
region on the night side.  The analytically computed values
(eq. \ref{eq:tn}) of the night-side temperature $T_n^a$ for each case
are also included in Table 1.

The second frame of Figure 1 shows the temperature distribution and
velocity pattern for Case 2, where the opacity is increased by a
scaling factor of $f_{\kappa} = 100$ over the standard interstellar
values.  This opacity enhancement relocates the photosphere, and the
opacity at the day-side photosphere $\kappa_d$ is increased by a
factor of 50.  The corresponding panel in Figure 4 shows the
temperature and velocity variations across the surface near the
photosphere. The result of the numerical simulation indicates $T_n ^n
= 230$ K. In accordance with equation (\ref{eq:tn}), $T_n ^a =200$ K
(see Table 1) which agrees well with the value of $T_n^n$.  Thus, the
temperature contrast between day and night sides is even larger than
that in Case 1, and the horizontal flow velocities are large, as a
consequence of the larger pressure gradient. Near the photosphere the
maximum horizontal velocity reaches $v_{\rm max} \sim 5$ km s$^{-1}$,
which is slightly supersonic relative to $c_d$ and highly supersonic
relative to $c_n$. Thus, at the anti-stellar point there is a shock,
which leads to enhancement of density and temperature by compressional
heating. This effect is clearly visible in the velocity and
temperature profiles (Fig. 4). The enhanced pressure behind the
shock leads to upward and downward flows near $ \theta =
180^\circ$. The downward flow feeds the return flow toward the hot
side.

As would be expected from equation (\ref{eq:pd}), photospheric
densities and pressures are considerably smaller in Case 2 than in
Case 1, since the photosphere lies at a higher level in the atmosphere
(see Fig.  1). The depth of the level where the horizontal flow
switches from positive to negative velocities also lies higher in Case
2.  Compared to Case 1, the region where the reverse flow leads to a
cold layer on the day side is larger in vertical extent and with a
lower minimum temperature of T=850 K. Note also that in this case the
reverse flow is still far below the photosphere but at a higher level
than in Case 1.

The temperature and velocity variations for Case 3, in which the
opacity is uniformly reduced by a factor 100 from the standard case,
are illustrated in the third frames of Figures 1 and 4. The
corresponding reduction in $\kappa_d$ is a factor of 25.  The level of
the photosphere is considerably deeper than in the previous cases, and
the pressures and densities are correspondingly enhanced in accordance
with equation (\ref{eq:pd}). The main day-to-night flow lies above the
photosphere and is much broader in vertical extent than in the
previous cases. The heated layer, which is denser than in Case 1,
carries a large amount of thermal energy from the day to the night
side.  The photospheric temperature on the day side is somewhat
reduced from the standard model in Case 1, and the photosphere extends
down to the cooler layer of the return flow. The reason for the lower
photospheric temperature is, however, not necessarily due to cooling
from the return flow, as explained below.  In Cases 1 and 2 the return
flow on the day side is heated by inward diffusion of radiation, while
in Cases 3 and 4 the upper layer of the return flow lies in  the
optically thin region and is directly heated.  The advective transport
of heat from the day to the night sides also increases the temperature
on the night side such that the temperature contrast between day and
night sides is much reduced compared with the previous cases, with a
higher average night-side temperature of about 850 K, in agreement
with equation (\ref{eq:tn}).  Consequently, the horizontal flow
velocities are reduced, with an average value on the night side of
$1.5 \times 10^5$ cm s$^{-1}$ as compared with $3.5 \times 10^5$ cm s$^{-1}$ in the
standard case.

On the night side the region where most of the radiative loss occurs
is above the photosphere near the terminator (Fig. 3).  But at larger
values of $x$, because of cooling, the scale height decreases and
material collapses to smaller radii. The flow toward the anti-stellar
point becomes increasingly concentrated near the photosphere.

In Table 2 we tabulate the gross heat flux $F_g = \rho_d v_d R_g
T_d/\mu$ at the photosphere on the day side.  The magnitude of $F_g$
is of order $10^{10}$ erg cm$^{-2}$ s$^{-1}$ in Case 3.  For Cases 3
and 4 $F_g$ is large compared to the radiative stellar heat flux,
which is of order $F_d = 3 \times 10^8$ erg cm$^{-2}$ s$^{-1}$. The
high value of $F_g$ is balanced by the flux in the return flow in the
high-density regions below the photosphere.

The trends seen in Case 3 continue in Case 4, where the opacity
reduction from the standard model is by a factor $10^3$. The
corresponding reduction in $\kappa_d$ is a factor of 250.  The
temperature and velocity patterns are shown in the fourth frames of
Figures 1 and 4. On the day side the photospheric pressures and
densities are about a factor 10 higher than in Case 3 (see Table 1).

The main day-night flow is again above the photosphere, which now
penetrates deeply into the layer of the cooler reverse flow. As the
temperature in the reverse flow is almost constant, the temperature
difference at the photosphere between the day and night side is
practically zero.  As the photosphere is actually in the return flow,
the velocity there is negative with a nearly constant value of 0.2
km/s.  The value of $T_d$ is reduced to 1030 K.

In addition to the back-flow effect, there is an additional reason why
there is a drop in $T_d$ compared with the previous cases. The total
incident luminosity from the stellar irradiation is $ (A/2) F_\ast$
where $F_\ast = L_\ast/4 \pi a^2$ is the stellar flux and $A/2$ is the
effective area of the day side.  The total radiation released from
both sides of the planet is $\approx (A/2)\sigma (T_d^4 + T_n^4)$. 
At  thermal equilibrium these two rates balance. Thus, if we define an
effective temperature $T_{\rm eff}$ such that $F_\ast = \sigma T_{\rm
eff}^4 = 1200$ K in our cases, then the temperature of the photosphere
at the day-side becomes

\begin{equation}
T_d = T_{\rm eff} \left(\frac{1}{1+T_n^4/T_d^4} \right)^{1/4}.
\end{equation}
Since $T_n\sim T_d$, $T_d$ is reduced by $2^{1/4} \sim 1.2$ compared
with $T_{\rm eff}$. Interestingly, the temperature in the return flow
at the photosphere happens to agree with this estimate.  In contrast
if $T_n << T_d$ as in Case 2, then $T_d \approx T_{\rm eff}$.

In Table 1 we compare the temperature obtained from our numerical
simulation ($T_n^n$) with that obtained from our analytic
approximation ($T_n ^a$).  The agreement between them is remarkable.
Our results can be compared with the estimates of
Showman \& Guillot (2002), who assume that grains are mostly depleted in
the planetary atmosphere and who adopted an equivalent opacity in their
calculation that is  comparable to or even lower than that which we
use in Case 4.  Based on a Newtonian local cooling approximation
(rather than diffusion through an opaque atmosphere), they estimate
the temperature difference $\Delta T_{dn} = T_d - T_n$ to be 500 K.
Our results do not agree with their conclusion, since we clearly
obtain $ T_d \approx T_n$ in Case 4.  If we use our cooling time
estimate from equation (\ref{eq:tauc}) for $\tau_{\rm rad}$ in
equation (\ref{eq:dn}), we do obtain only a 10\% difference between
$T_d$ and $T_n$. The difference must lie in their alternate approach
to the calculation of $\tau_{\rm rad}$. These results clearly indicate
that 1) the atmospheric dynamics are driven by the radiation transfer
process, and 2) the local Newtonian cooling prescriptions adopted in
the previous investigations are oversimplified.

\section{Energy deposition and transport}

\subsection{Shear and kinetic heating of interior regions}

The total heating rate ${\dot E}_d$, as well as the gross and net
advective heat fluxes near the photosphere ${\dot E}_{g}$ and ${\dot
E}_{a}$, are much larger than the intrinsic luminosity of a
few-Gyr-old Jupiter-mass planet.  The Kelvin-Helmholtz contraction and
internal cooling of our standard model planet give a flux of heat loss
$F_{\rm p} \approx 3000$ erg cm$^{-2}$ s$^{-1}$, which is five orders
of magnitude smaller than $F_d \approx 3 \times 10^8$ erg cm$^{-2}$
s$^{-1}$. Guillot \& Showman (2002) suggest that a modest amount of
dissipation of the energy associated with the advective flow, if deep
enough in the atmosphere, could produce sufficient heating and
mechanical work to expand the atmosphere.  The circulation pattern
which has been described in previous sections produces a shear, which
is most important at the interface between the day-to-night flow and
the return flow.  Showman \& Guillot (2002) suggest that a shear layer
of this type could provide a kinetic heating source through the
effects of Kelvin-Helmholtz instabilities.  But they require that the
major dissipation take place at lower levels in the atmosphere than
the shear layer evident in Figure 1.

As a first approximation of this heating effect, we assume 
an effective viscosity $\nu$, which leads to 
a rate of energy dissipation per unit volume of 
\begin{equation}
{\dot e}_k = \rho \nu \left({d v_x \over dy} \right)^2.
\end{equation}
Integrating over the entire volume $V$  of the atmosphere, the total
kinetic energy dissipation is $\dot E_k  = \int \dot e_k dV $
 (the $x-$ velocity 
gradient clearly dominates over that in $y$).

Although we do not include the effect of viscous drag in the momentum
equation which we solved numerically, we can compute the distribution
of $\dot e_k$ from our results.  As in typical astrophysical
situations, molecular viscosity is too small to make any significant
contribution.  The flow near the interface layer does in fact lead to
a Richardson number less than the critical value of 1/4, which would
induce a shearing instability (Showman \& Guillot 2002). However,
modification of the stability criterion is needed from the traditional
Boussinesq approximation, because both the stabilizing buoyancy effect
and the destabilizing shear are modified by both heating and cooling
efficiencies (Garaud \& Lin 2003; Jones \& Lin 2003).  If the shear
flow were unstable, we could determine a viscosity with an {\it ad
hoc} $\alpha$ prescription (Shakura \& Sunyaev 1973), in which
\begin{equation}
\nu = \alpha c_s h_s \simeq \alpha c_s^3 / g, 
\end{equation}
where $c_s$ is the local magnitude of the  sound speed and $h_s$ is the
local pressure scale height.  

For Cases 1 and 3 we obtain the $\dot e_k$ distribution shown in
Figure 5 (for $\alpha =1$).  This figure shows that for this
particular prescription of determining the dissipation, the kinetic
dissipation rate has very little effect on the energy flow at high
pressures, because the shear at lower levels in the atmosphere is
small.  The most intense dissipation occurs mostly on the night side
in the vicinity of the shear layer.

In Figure 6 we plot the quantity $ \alpha^{-1} F_k$, where $F_k$ is
defined as the total energy dissipation rate in the grid per unit
surface area, integrated over all $x$ and from the upper edge of the
grid down to some cutoff pressure $P$.  The results show that the
quantity $F_k$ saturates to a value $F_{\rm k,tot}$ in the deeper
layers, indicating that the major contributions occur at relatively
low optical depth, in regions where $d F_k/dy$ is at a maximum and the
local pressure $P_m$ (see Table 2) is in the range $~10^{-3}$ to 0.1
bar.  The total amount of energy dissipated per unit area, indicated
by the horizontal parts of the curves, increases roughly as $F_{\rm
k,tot} \propto 1/\sqrt{\kappa}$.

The total integrated values of $\alpha^{-1} F_k$ (asymptotic value in
the limit of high $P$), corresponding to the highest values on the
curves in Figure 6, are shown in Table 2.  These values should be
compared with the incident radiation flux $F_d \approx 3 \times 10^8$
erg cm$^{-2}$ s$^{-1}$. For the high opacity Cases 1 and 2 the
integrated flux $F_k$ can be quite significant, if $\alpha$ is greater
than of order $10^{-2} - 1$.  In the low opacity cases the integrated
$F_{\rm k,tot}$ is smaller than $F_d$ only if $\alpha < \approx 5
\times 10^{-3}$.

If the kinetic energy associated with the shear is dissipated very
close to the surface, it would be radiated without significantly
modifying the planet's envelope.  If, on the other hand, 
 a significant fraction ($\sim
10^{-2}$) of the stellar irradiative flux is dissipated in a region
with pressure between 10 and 100 bar, the envelope may adjust to a new
hydrostatic equilibrium by expanding its radius (Guillot \& Showman
2002).  From Figure 6 we find that if $\alpha \approx
10^{-2}$, more than $99\%$ of the total incident stellar irradiative
flux is dissipated at a pressure which is well below 1 bar.  Little or
no expansion is expected.

The results in Figure 6 also justify our neglect of the viscous
effects in the numerical treatment of the momentum and the energy
equations for modest values of $\alpha$, because much of the energy
dissipation that occurs in the optically thin zones is radiated
locally.  However, in the high viscosity ($\alpha$) limit where
\begin{equation}
{\dot F}_k  \ge \dot F_d,
\end{equation}
the viscous momentum transfer and energy dissipation would reduce the
velocity gradients until a new equilibrium is established whereby ${\dot
E}_k < \dot E_a$.  In this regard, we tabulate in Table 2 a
quantity
\begin{equation}
f_k \equiv F_k (y_1)/ \alpha F_\ast, 
\end{equation}
where $F_k (y_1)$ is the energy dissipation flux between $y_{max}$ and
the $y$-level where $P = 1$ bar, where most of the energy is
dissipated. For all cases $f_k > 1$ for $\alpha=1$. But $f_k$ should be 
less than unity, which means 
 $\alpha$ is required to be
less than $\sim 10^{-2}$.  The results in Table 2 indicate that $f_k$
increases as the opacity drops. This dependence arises naturally since
the pressure at the photosphere increases as $\kappa_d$ drops.
Although stellar irradiation is able reach deep down in the
low-opacity Cases 3 and 4, the smaller temperature differences
between the day and night sides of the planet limits the magnitude of
the flow speed and therefore that of the shear.

Based on these results, it is worthwhile to further examine the effect
of energy dissipation associated with the circulation by including the
viscous transport and energy dissipation contributions in the
governing equations.  In our calculations we also neglected rotation,
especially the Coriolis effect.  In principle, this process can induce
the circulation current to mix to a different depth.  We will check
the radial extent of this mixing in the future.

\subsection{Heat transfer}

In the absence of any significant surface irradiation the envelope of
a mature gas giant planet is nearly fully convective.  In our
particular spherically symmetric planetary structure models the
surface irradiation suppresses convection down to $\sim 0.9 R_p$
(Bodenheimer et al. 2001).  The results in Figure 1 indicate that the
convection in the planetary envelope is suppressed at least on the day
side.  The extent of the radiative zone affects the planet's structure
and evolution in four ways.  1) The heat diffusion flux through the
radiative zone may lead to a reduction in the planet's cooling rate.
2) A stable thermal stratification may also suppress the onset of
shearing instability associated with the circulation flow.  3) The
suppression of convection may enhance the rate of grain sedimentation
and modify the magnitude of the opacity.  4)  The tidal disturbance can
lead to $g$ mode excitation (Lubow et al. 1997) in radiative zones, 
whereas Hough modes may be excited in convective zones (Ogilvie \& Lin
2003).

We plot in Figure 7 the distribution of the local normalized Brunt--V\"ais\"al\"a
frequency where $N_n^2 \equiv R_p^3 N^2 / G M_p$ for
Cases 1 and 3 (with standard and reduced opacities). The magnitude is
determined from
\begin{equation}
  N^2={{1}\over{\rho}}{{dp}\over{dr}}
  \left({{d\ln\rho}\over{dr}}-{{1}\over{\gamma}}{{d\ln p}\over{dr}}\right),
\end{equation}
where $\gamma$ is the adiabatic exponent. Positive and negative values
of $N_n^2$ represent the radiative and convective regions
respectively.  In both cases the day side is radiative throughout the
computational domain.  Note that the inner boundary condition we have
adopted is matched to the radiative region deduced from the 1-D
spherical calculation, which includes the effect of stellar
irradiation.  The magnitude of $N_n^2$ is maximum near the photosphere
where the stellar irradiation is mostly absorbed.

On the night side the surface region becomes convective near the
photosphere in Case 1.  The extent of the convective region appears to
be confined.  Below it there is a relatively intensive radiative
layer which is also confined.  In Case 3 thermal energy transport by
the circulation current does not alter the radiative state on the
night side throughout the entire computational domain.  The
distribution of $N_n^2$ is independent of the longitude ($x$) in the
deeper regions in Case 1.  A slight longitudinal variation in $N_n^2$
near the inner boundary in Case 3 may be attributed to the fact that
thermal equilibrium has not completely been established for this low
opacity case.

The existence of a  radiative surface layer supports the suggestion that
g-mode oscillations may be excited just above its interface with the
planet's convective envelope and dissipated through radiative or
nonlinear damping (Lubow, Tout, \& Livio 1997).  Outwardly propagating
Hough waves may also be excited within this radiative layer and be
damped in the atmosphere (Ogilvie \& Lin 2003).  This process may play
a dominant role in driving the short-period extrasolar planets toward
a state of synchronous rotation.

\section{Summary and discussion}

In this paper  we consider the circulation arising from  stellar
irradiation in the atmospheres of synchronously-spinning short-period
planets around nearby stars.
Our results confirm early suggestions that one-sided heating can lead
to circulation flow from the day to the night sides of these planets
(Showman \& Guillot 2002; Cho et al. 2003).  This circulation current
carries with it considerable thermal energy content,  which provides a
heating source for the planets' night sides.  At the night side of the
planets where the circulation flow converges, the cool and dense gas
submerges and induces a reverse flow.  The returning current occurs
well below the photosphere. The returning current is generally cooler
than its ambient gas.  Consequently, radiation diffusion into this
layer is enhanced.  The returning current eventually converges on the
sub-stellar point where the diverging flow near the surface leads to a
local reduction of density and pressure. The resurfaced current
completes a circulation pattern with the excess heat transported
advectively to the surface and radiated at the photosphere of both day
and night sides of the planet.  This flow pattern indicates that the
surface irradiation on a synchronously spinning planet may not
suppress the cooling of its interior as much as suggested by Burrows
 et al. (2000, 2003) on the basis of a one-dimensional model.  However,
a more rigorous follow-up study is needed.

In contrast to previous investigations, we take into account the
effect of radiative diffusion in the planetary atmosphere and
envelope.  We show that the night side temperature is a sensitive
function of the atmospheric opacity.  If the grain content in the
atmosphere were comparable to that in the interstellar medium, only a
shallow layer in the planetary atmosphere would be exposed to the
stellar irradiation. This limited supply of heat to the night side
would lead to a relatively low effective temperature there.  The
relatively large azimuthal temperature and pressure differential along an 
equipotential surface would induce the flow to become transonic.  If
the shear flow induced by the circulation pattern is unstable and the
resultant turbulence is efficient in inducing mixing between the
layers, the dissipation of energy can be modest. However, most of this
energy dissipation occurs near the photosphere such that it is likely
to be rapidly radiated away.

If the grains were depleted, the stellar irradiation would be able to
penetrate deeply into the planet's atmosphere.  The circulation
current would carry much greater mass and thermal energy flux than in
the standard Case 1. Since only a portion of the incident stellar
thermal energy flux is reprocessed, the day-side effective temperature
for the low-opacity cases is lower than that for the normal and
high-opacity cases.  In contrast, the night side receives the heating
induced by advective circulation, so the surface temperature is
relatively large.  Both of these effects reduce the temperature
differential between the day and night side.  The reduced circulation
speed implies that less energy would be dissipated.  Although the
photosphere extends deeply into the planet's atmosphere, the
dissipation of the circulation flow would occur below the photosphere.
For the reduced-opacity Cases 3 and 4, the energy release, as a
consequence of this dissipation, occurs well below the surface.  Our
estimated rate of energy dissipation does not appear to be adequate to
account for the inflation  of the short-period planet around HD209458 to
its presently observed value (Brown et al. 2001).

The above summary clearly indicates that the radiative properties and
the dynamical structure of the atmosphere of close-in planets depend
sensitively on the presence and depletion of grains.  We have not
considered the dynamical effect of the circulation on the grain
evolution in the planetary atmosphere.  A preliminary estimate
suggests that the characteristic time scale of the circulation is
comparable to the sedimentation time scale of relevant size grains in
the planetary envelope.  In a subsequent paper we shall examine the
self-consistent evolution of the grains in such an atmosphere.

We also neglected the effect of rotation which introduces a Coriolis
force to the flow.  This effect may lead to latitudinal circulations
as well as vorticity stretching.  It may suppress heat flow from the
equator to the poles as well as break the hemispheric longitudinal
symmetry.  It could also induce the circulation to penetrate farther
into the planet's interior and to couple the flow in the azimuthal and
radial directions.  Finally, it could lead to instabilities to enhance
the dissipation of the shear associated with the circulation flow.
These issues need to be examined in three dimensions with a proper
treatment of the radiation transfer, a problem which will also be
considered in a future study.

\acknowledgements We wish to thank G. Novak for useful conversation
and valuable comments. This work is supported by NSF and NASA through
grants AST-9987417 and NCC2-5418. This work was supported in part by
the European Community's Human Potential Programme under contract
HPRN-CT-2002-00308,PLANETS.

\clearpage

\vskip 0.2 in
TABLE 1
\vskip 0.2 in

\begin{tabular}{c c c c c c c c c c} 
\hline \hline & & & & & & & & &\\ 
Case& $T_n^n$ & $T_n^a$ & $\rho_n$ & $P_n$ &$\kappa_d$& $T_d$ & $\rho_d$ 
& $P_d$ & $v_{max} $ \\
 & & & & & & & & & \\ \hline  & & & & & & & & & \\ 
1 & 480 & 490 & $3 \times 10^{-8}$ & $6 \times 10^{-4}$ & 1 & 1200 & $3
\times 10^{-8}$ &$1.2 \times 10^{-3}$& $3.5 \times 10^5$ \\ 
& & & & & & &  & & \\  & & & & & & & & & \\ 
2 & 230 & 200 & $3 \times 10^{-9}$ & $4 \times 10^{-5}$ & 50 & 1200 &
$4 \times 10^{-10}$ & $2 \times 10^{-5}$ & $5 \times 10^5$ \\ 
& & & & & & & & & \\ & & & & & & & & & \\ 
3 & 850 & 840 & $7 \times 10^{-7}$ & $3 \times 10^{-2}$ & 0.04 & 1100
& $5 \times 10^{-7}$ &$2 \times 10^{-2}$& $1.2 \times 10^5$ \\
& & & & & & & & & \\ & & & & & & & & & \\ 
4 & 990 & 960 &$6 \times 10^{-6}$ & 0.25 & 0.004 & 1050 & $4 \times
10^{-6}$ & 0.17 & $ < 2 \times 10^4$ \\ 
& & & & & & & & & \\ & & & & & & & & & \\  \hline
\end{tabular}

\clearpage
\vskip 0.2 in
TABLE 2
\vskip 0.2 in

\begin{tabular}{c c c c c c c } 
\hline \hline & & & & & &  \\ 
Case& $F_g$ & $F_d$ & $F_n$ & $\alpha^{-1} F_k$  &  $P_m$ 
& $f_k$ \\
& & & & & &  \\ \hline & & & & & &  \\ 
1 & $9 \times 10^8$ & $3 \times 10^8$ & $3 \times 10^{6}$ & $3.3 \times 10^{9}$ & 
 $2 \times 10^{-3}$ & 11 \\ 
& & & & & &  \\ & & & & & &  \\ 
2 & $4 \times 10^7$ & $3 \times 10^8$ & $1.6 \times 10^{5}$ 
& $9.6 \times 10^{8}$ & $1 \times 10^{-3}$  &  
 3.2 \\ 
& & & & & &  \\ & & & &  & & \\ 
3 &  $6 \times 10^9$ & $3 \times 10^8$ & $ 3 \times 10^{7}$ & $2.7 \times 10^{10}$ 
&  $3 \times 10^{-2}$ & 90 \\ 
& & & & & &  \\ & & & & & &  \\ 
4 & $8 \times 10^9$ & $3 \times 10^8$ & $5.4 \times 10^{7}$ & $5 \times
10^{10}$ &  0.1 & 167 \\ 
& & & & & &  \\ & & & & & &  \\  \hline
\end{tabular}

\begin{figure} 
\plotone{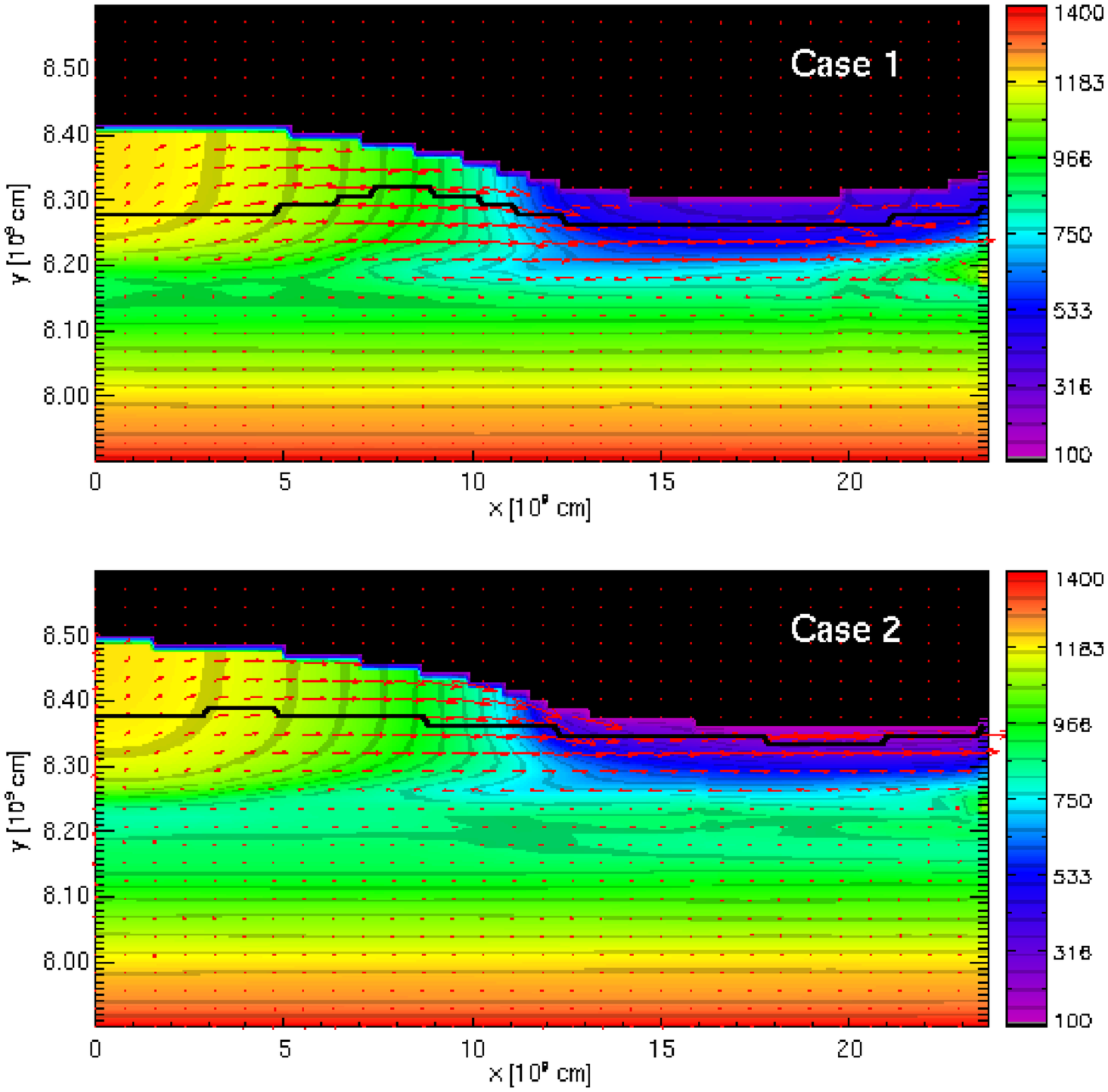}

{\bf Fig. 1a.-- }  Colors represent  temperatures in the ($x,y$) plane for Cases 1 and 2 
when they have reached steady state. 
Color scale runs from 1400 K (red) to 100 K (violet). The black line
corresponds to the photosphere ($\tau = 2/3$).  Arrows correspond 
to velocity vectors, with length proportional to speed and with a
maximum of 3 km/s. 
\end{figure}

\begin{figure} 
\plotone{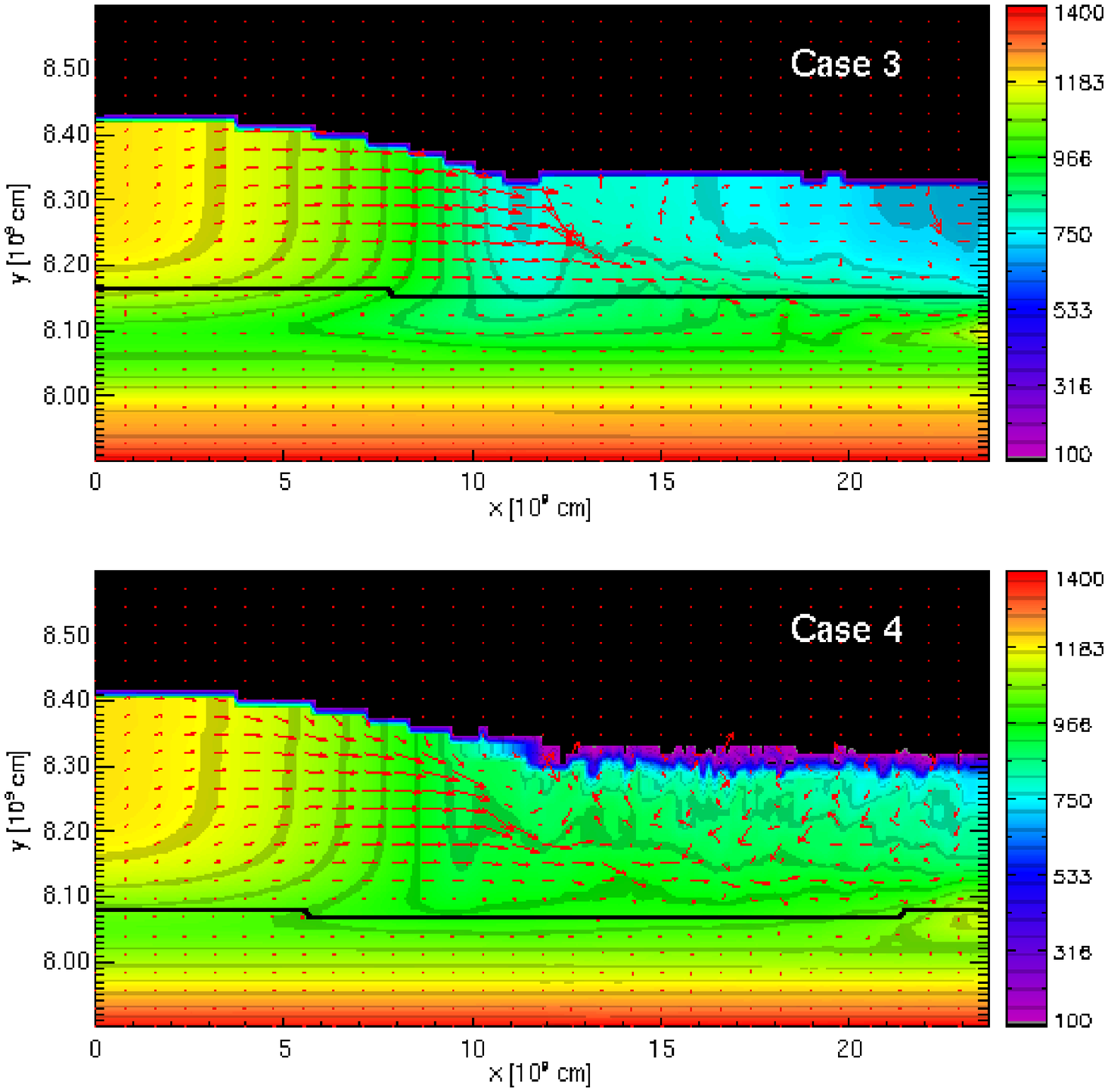}

{\bf Fig. 1b.-- } Same as Fig. 1a except that Cases 3 and 4 are plotted. 
\end{figure}

\begin{figure} 
\plotone{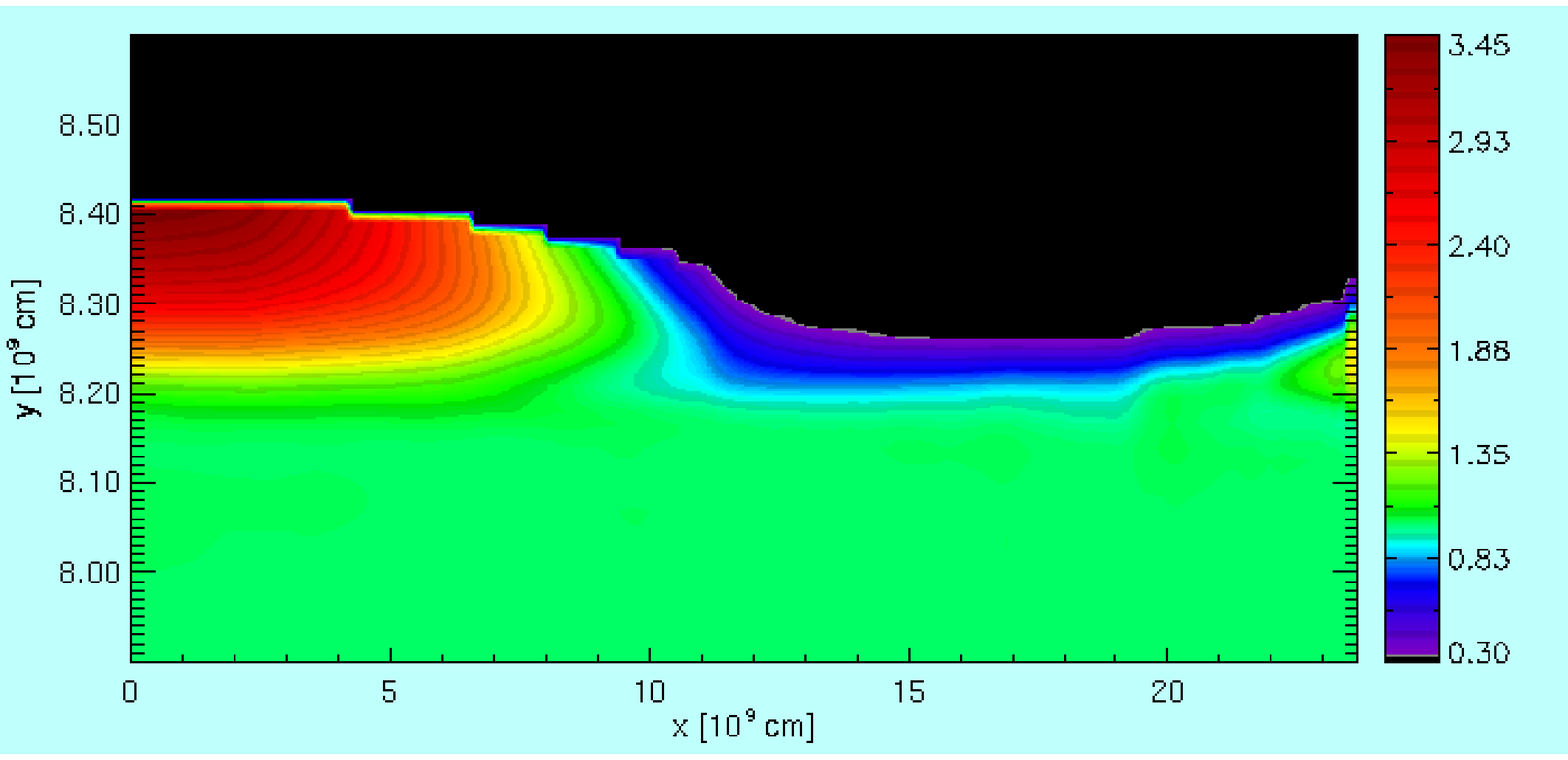}

{\bf Fig. 2.-- } {The quantity $f_p$,  defined as the ratio of the local pressure to the
horizontally averaged pressure at the given value of $y$, 
is plotted in the $(x,y)$ plane for Case 1 after the calculation has reached
steady state. {\it Red:} high-pressure regions; 
{\it Blue to violet:} low-pressure regions; {\it Green:} regions where there is
no significant local deviation from the mean pressure at the  given value of $y$.}
\end{figure}

\begin{figure} 
\plotone{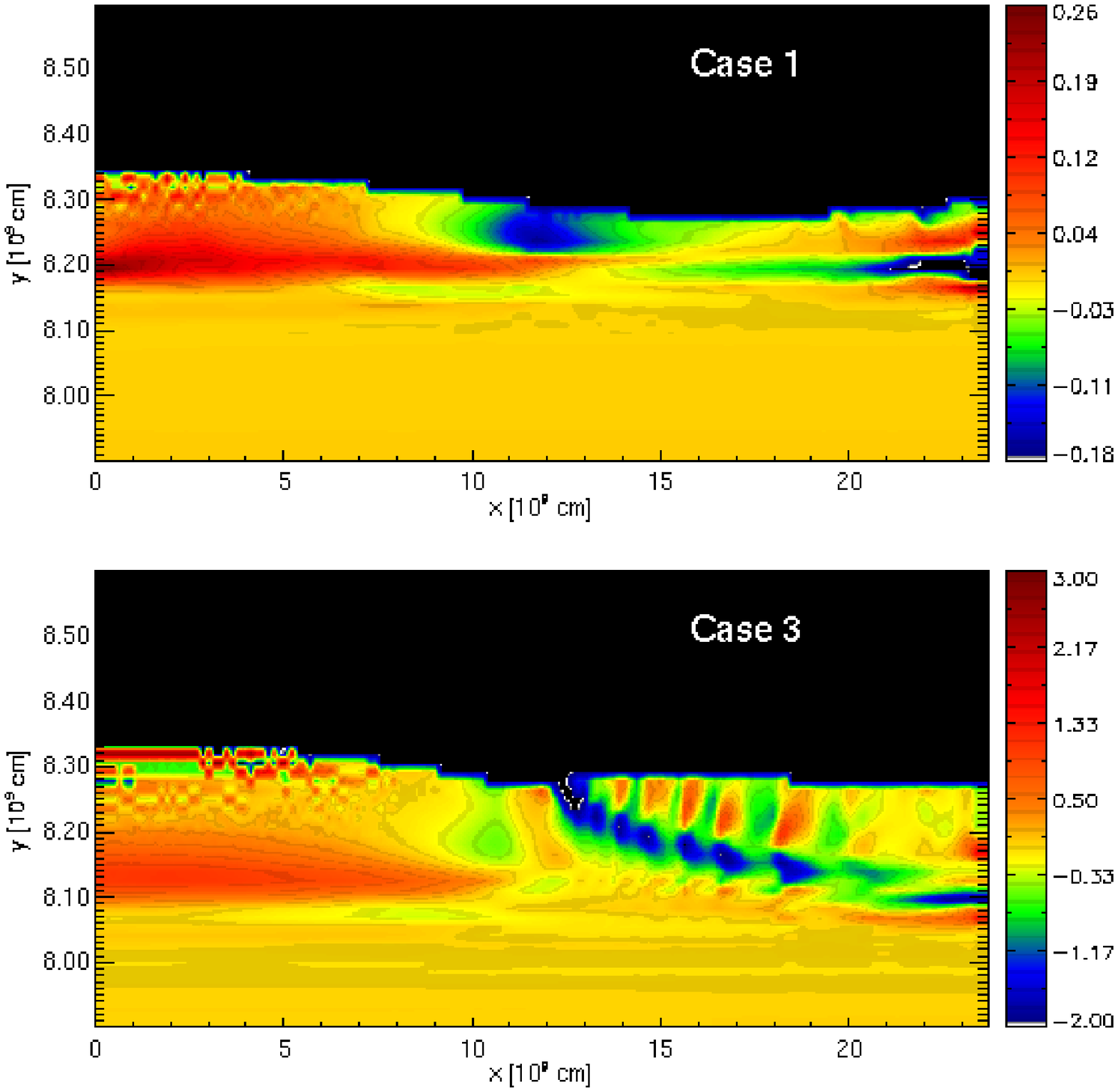}

{\bf Fig. 3.-- } {The local rate of change of internal energy per unit volume 
caused by radiation transfer (right-hand side of eq. \ref{eq:energy})
when the model has reached steady state.
Colors represent regions in the $(x,y)$ plane which are strongly heated
({\it brown to red }) and which are strongly cooled ({\it blue}).
{\it Upper panel}:  Case 1;  {\it lower panel}:  Case 3.}
\end{figure}
                  
\begin{figure} 
\plotone{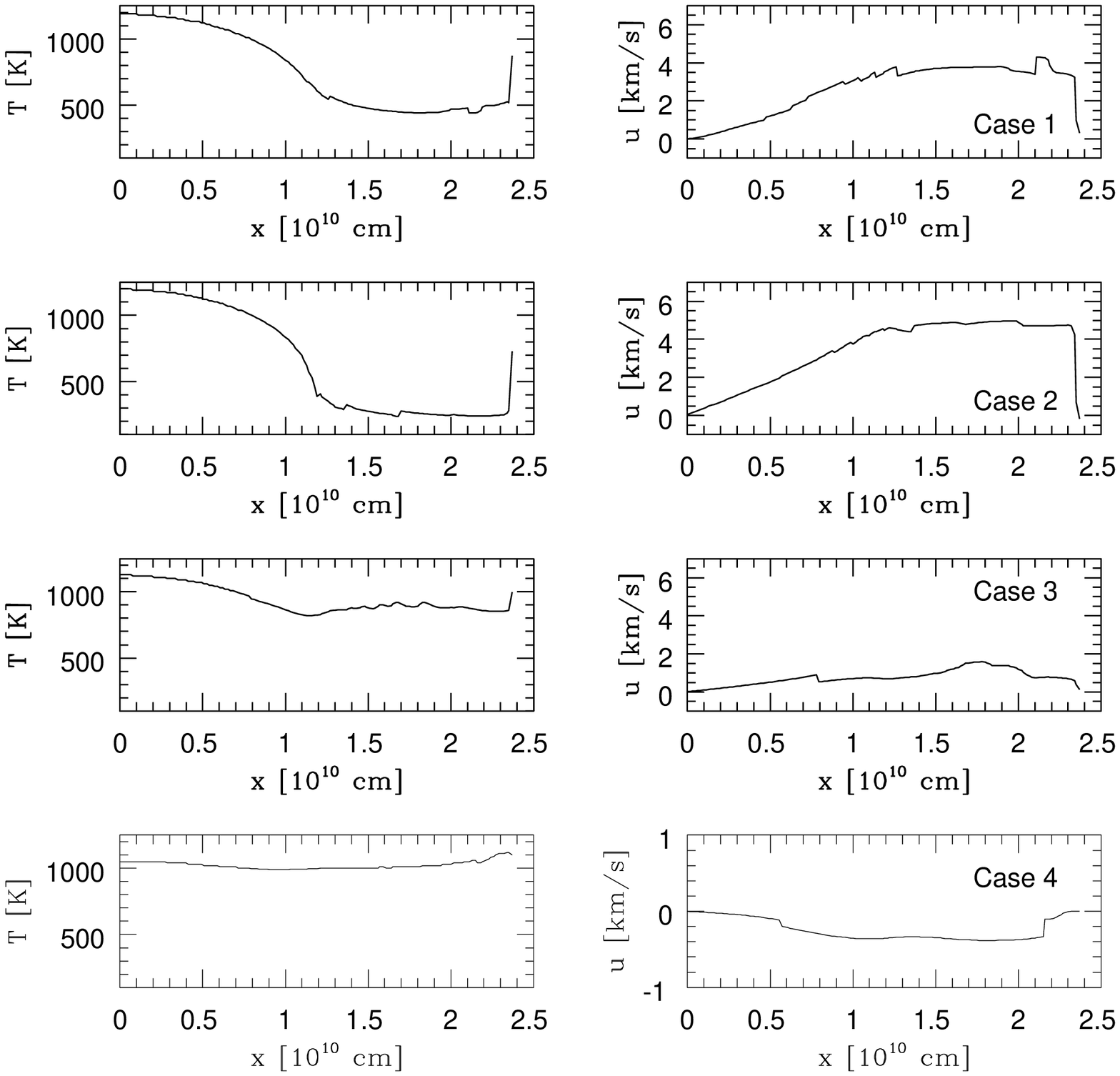}

{\bf Fig. 4.-- } {The left frames  give the temperature as a  function of $x$
at the photosphere for the steady-state model.  The right frames give the velocity in
the $x$-direction as a function of $x$, also at the photosphere.  {\it Top to 
bottom:} Cases 1, 2, 3, and 4. }
\end{figure}

\begin{figure} 
\plotone{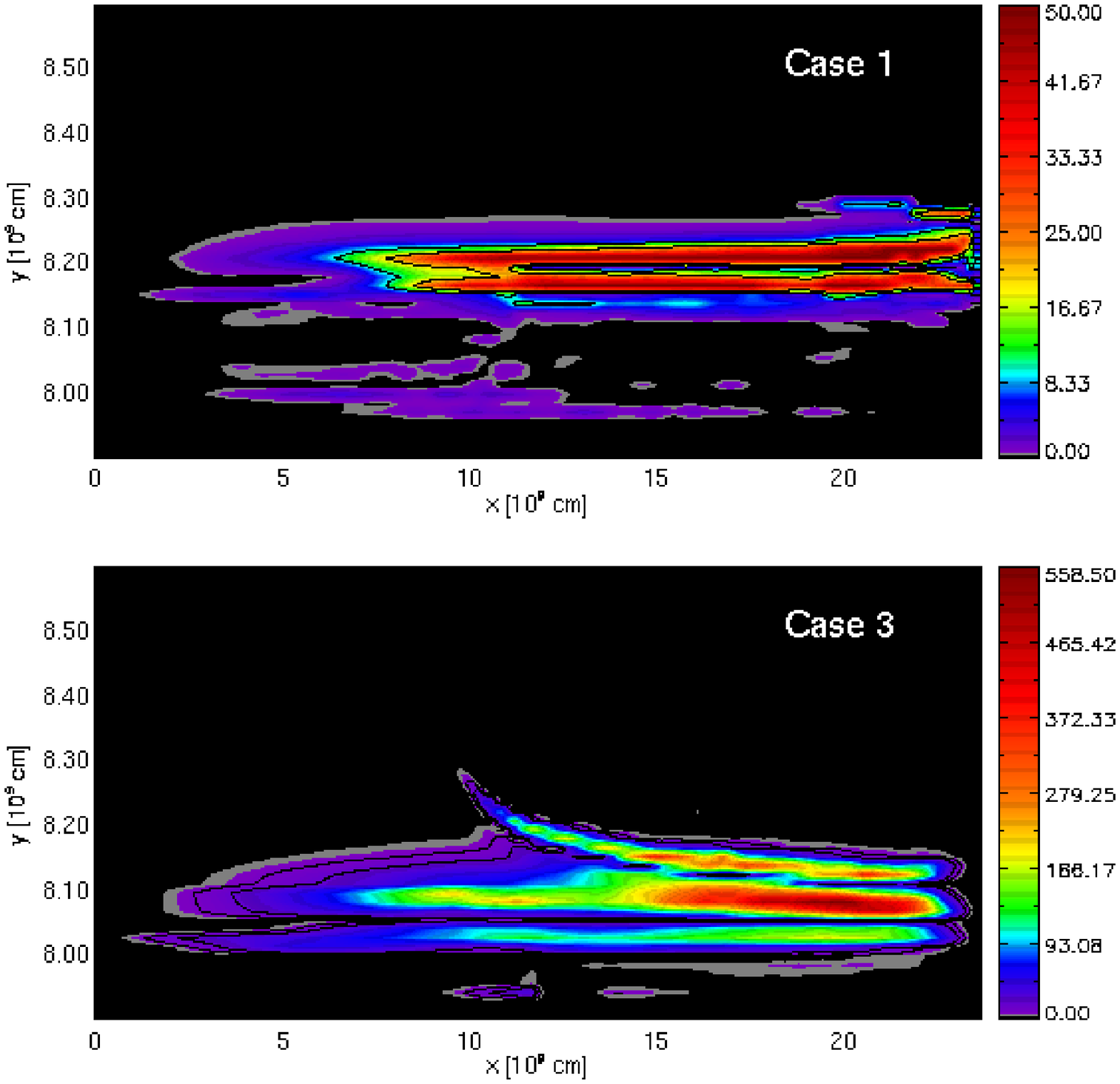}

{\bf Fig. 5.-- } {The local rate of viscous heating is plotted in the $(x,y)$ plane.
{\it Upper panel}:  Case 1;  {\it lower panel}:  Case 3.}
\end{figure}

\begin{figure}
\plotone{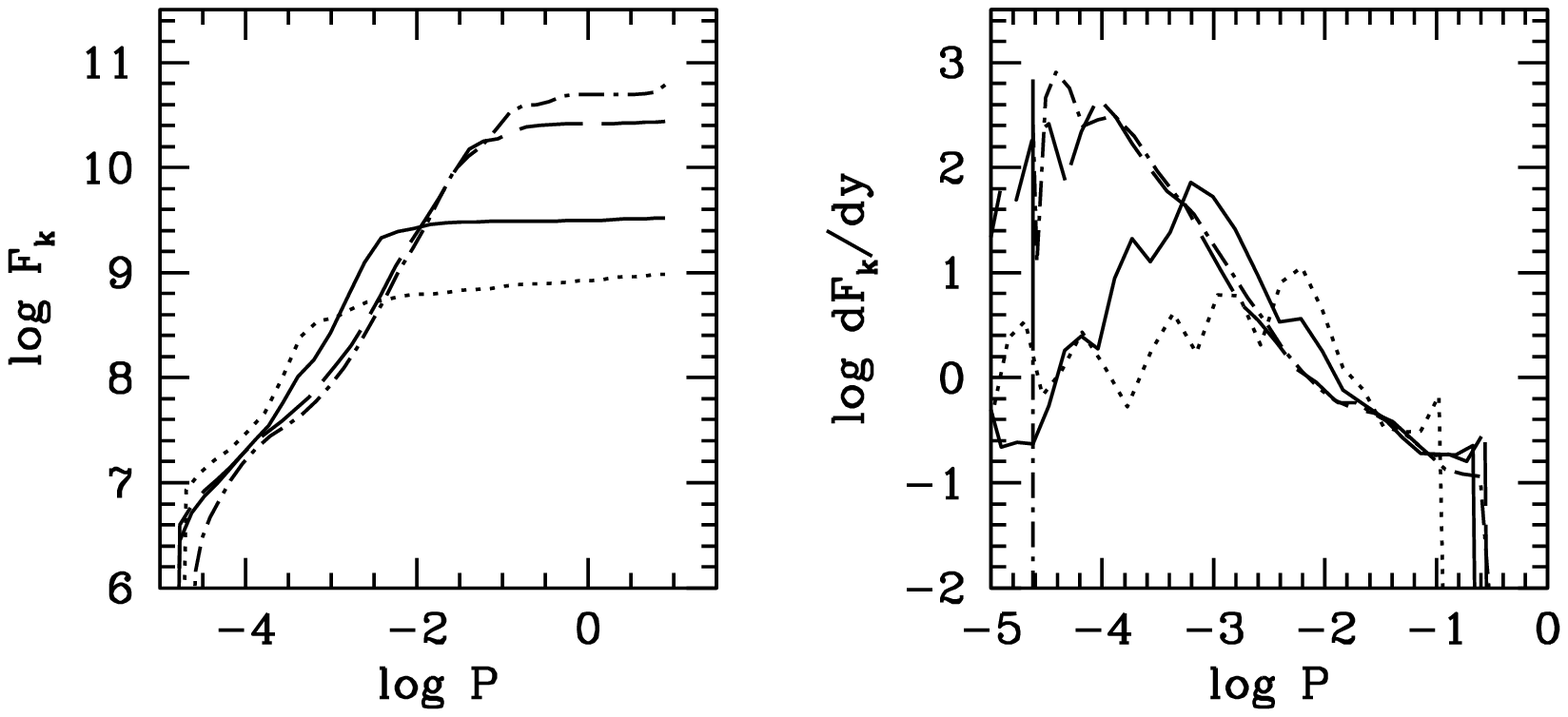}

{\bf Fig. 6.-- } {{\it Left frame}: The  quantity $\alpha^{-1} F_k$ is plotted, 
where $ \alpha$ is the viscosity parameter and $F_k$ is 
the rate of viscous dissipation per unit area,
 integrated downwards to the level where the pressure is $P$. 
{\it Solid line:} Case 1; {\it dotted line:} Case 2;  {\it dashed line:} 
Case 3; {\it dash-dotted line:} Case 4. {\it Right frame:} the derivative of
the function plotted in the left frame. }
\end{figure}

\begin{figure}
\plotone{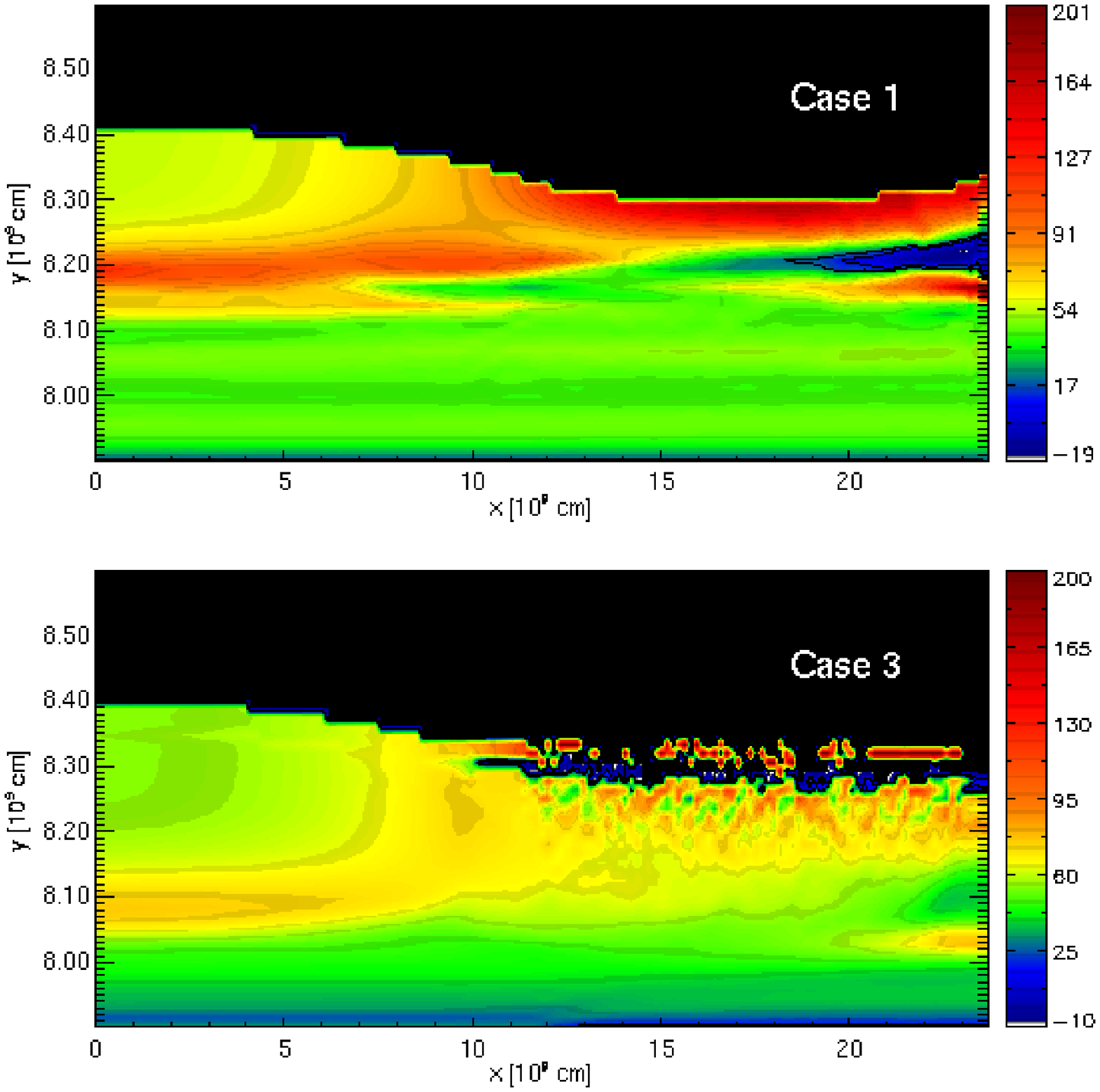}

{\bf Fig. 7.--}  The normalized squared Brunt-V\"ais\"al\"a frequency $N_n^2$
 is plotted in the $(x,y)$ plane. 
{\it Upper panel}:  Case 1;  {\it lower panel}:  Case 3.
\end{figure}

\end{document}